\documentclass[a4paper,superscriptaddress,notitlepage,12pt,tightenlines,amsmath,amssymb,pra]{revtex4-2}

\usepackage{graphicx}
\usepackage{amsmath,amssymb,setspace}
\usepackage{newtxtext}
\usepackage{newtxmath}
\usepackage{verbatim}
\usepackage[x11names]{xcolor}
\usepackage[unicode,colorlinks,citecolor=Blue3,linkcolor=Blue3,bookmarks=true]{hyperref}
\DeclareMathOperator*{\ch}{cosh}
\DeclareMathOperator*{\sh}{sinh}

\newcommand{\mean}[1]{\langle{#1}\rangle}
\newcommand{\svector}[2]{\begin{pmatrix}#1 \\ #2 \end{pmatrix}}
\newcommand{\smatrix}[4]{\begin{pmatrix}#1 & #2 \\ #3 & #4\end{pmatrix}}
\newcommand{\ds}{}
\newcommand{\dr}{}

\begin{document}

\title{Amplified quantum non-demolition measurements of optical quadratures using quadratic nonlinearity}

\author{D. I. Salykina}%
\email{koil257@mail.ru}
\affiliation{Russian Quantum Center, Skolkovo IC, Bolshoy Bulvar 30, bld.\ 1, Moscow, 121205, Russia}
\affiliation{Faculty of Physics, M.V. Lomonosov Moscow State University, Leninskie Gory 1, Moscow 119991, Russia}

\author{V. S. Liamin}%
\affiliation{Russian Quantum Center, Skolkovo IC, Bolshoy Bulvar 30, bld.\ 1, Moscow, 121205, Russia}

\author{P. R. Sharapova}
\affiliation{Department of Physics, Paderborn University, Warburger Str. 100, D-33098 Paderborn, Germany}

\author{F. Ya. Khalili}
\email{farit.khalili@gmail.com}
\affiliation{Russian Quantum Center, Skolkovo IC, Bolshoy Bulvar 30, bld.\ 1, Moscow, 121205, Russia}

\begin{abstract}
  Quantum non-demolition (QND) measurement is a special  technique that allows to evade quantum back-action. In this paper, we propose a new QND measurement scheme of the optical field quadratures based on the non-degenerate optical parametric amplifier. We show that for a proper set of parameters, this scheme can realize a new type of QND measurement, where the quadrature of interest is amplified, but still does not subject to any back action.
\end{abstract}

\maketitle


\section{Introduction}



According to the von Neumann's reduction postulate, quantum mechanics allows measuring one given observable without perturbing it \cite{von2018mathematical, bohm1989quantum}. In late 1970s, the concept of the  Quantum Non-Demolition (QND) measurement \cite{braginskii1977quantum, Braginsky1980, braginsky1995quantum, Braginsky1996} implementing the reduction postulate was formulated. It was shown, in particular, that a {\it sufficient} (not necessary) condition for a QND measurement is the commutativity of the measured observable with both the Hamiltonian of the measured system and the Hamiltonian describing the interaction of this system with the measuring device.

In Refs \cite{braginskii1981non, Milburn1983}, it was proposed to use the Kerr ($\chi^{(3)}$) optical nonlinearity to implement the QND measurement of the optical photon number. Later, several proof-of-principle experiments based on this idea were performed, see the review \cite{grangier1998quantum}. However,  because of the weakness of the Kerr nonlinearity in highly transparent optical media, the sensitivity only slightly better than that of {\dr Standard Quantum Limit (SQL) \cite{braginsky1995quantum, Braginsky1996}} was achieved in the mentioned experiments. This problem can be alleviated by using the high-$Q$ optical microresonators \cite{strekalov2016nonlinear}, which combine very low optical losses with a high concentration of the optical energy, and was analyzed theoretically in Refs.\,\cite{Balybin2022, balybin2023improving, salykina2025intracavity}. {\dr An interesting alternative strategy allowing to circumvent the need for strong Kerr nonlinearity by using the second-order ($\chi^{(2)}$) one to mimic an effective $\chi^{(3)}$ QND type interaction was proposed in Ref.\,\cite{yanagimoto2023quantum}.}

\begin{figure}
\includegraphics[width=0.7\textwidth]{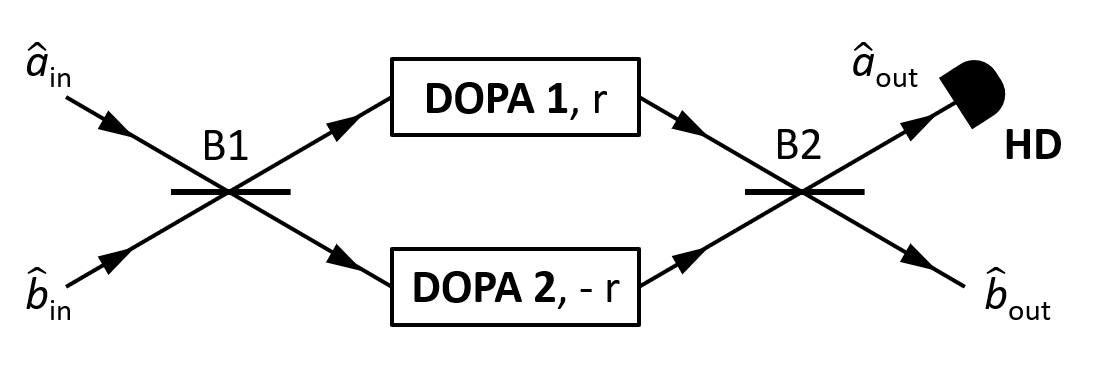}
\caption{\label{fig1} The QND scheme for quadrature measurement proposed in Ref.\,\cite{yurke1985optical}. DOPA -- degenerate optical parametric amplifier, ${\rm B}_1$, ${\rm B_2}$ -- beam splitters.
}\label{fig:DOPAs}
\end{figure}

Another promising approach to QND measurements is the measurement of a quadrature amplitude of light {\dr by means of the Optical Parametric Amplifier (OPA) type interaction. It can be implemented by means of both second-order ($\chi^{(2)}$) and third-order ($\chi^{(3)}$) nonlinearities using, respectively, the three-wave mixing and the four-wave mixing processes.} Possible implementations of this type of measurement, {\dr shown in Fig.\,\ref{fig:DOPAs},} were proposed  in Ref.\,\cite{yurke1985optical} and experimentally realized in Ref.\,\cite{la1989back}. Here again, only marginal sensitivity improvement ($\approx0.6\, {\rm dB}$) compared to the SQL was achieved. However, it is hoped that with this type of QND measurements, much better sensitivity could be obtained using the high-$Q$ optical microresonators fabricated from media with the $\chi^{(2)}$ nonlinearity.

At the same time, the scheme proposed in Ref.\,\cite{yurke1985optical} has one significant disadvantage. It requires two degenerate optical parametric amplifiers (OPAs) with strictly anti-balanced gain factors. This requirement could hinder the scheme implementation based on the microresonators.

In the current work, we present a more practical version of the optical quadrature amplitude QND measurement based on $\chi^{(2)}$ nonlinearity that uses only a single non-degenerate optical parametric amplifier (NOPA). It is interesting to note that the NOPA interaction Hamiltonian alone does not satisfy the above mentioned sufficient condition of QND. This is true also in the case of the initial DOPAs based scheme of Ref.\,\cite{yurke1985optical}. Therefore, in both cases, the additional passive optical elements (two beamsplitters) should be used to implement the QND type interaction.

Moreover, we show that combining the NOPA with a single beamsplitter, it is possible to implement a novel type of quantum measurement which we propose to call the amplified Quantum Non-Demolition (AQND) measurement. This scheme supplement the QND-type measurement by noiseless amplification of the measured observable, improving, in particular, its robustness to optical losses.

The paper is organized as follows. In the introductory Section \ref{sec:intro}, we discuss the general features of the $\chi^{(2)}$ based QND measurement of optical quadrature amplitude, using as an example the scheme of Ref.\,\cite{yurke1985optical}. In Sec.\,\ref{sec:schemes} we introduce the NOPA-based QND and amplified QND {\dr schemes}. In Sec.\,\ref{sec:losses}, we {\dr estimate} the sensitivity of the proposed schemes, taking into account the optical losses. In Sec.\,\ref{III} we {\dr discuss the prospects of experimental implementation of proposed schemes and} summarize the main results of this paper.

\section{The idea of the $\chi^{(2)}$-based QND measurement}\label{sec:intro}

Let us introduce the cosine (amplitude) and sine (phase) quadratures of the optical field as follows:
\begin{gather}\label{quadratures}
  \hat a^c = \frac{\hat a + \hat a^{\dagger}}{\sqrt 2},
  \quad \hat a^s = \frac{\hat a - \hat a^{\dagger}}{i\sqrt 2},
\end{gather}
where $\hat{a}$, $\hat{a}^\dag$ are the corresponding annihilation and creation operators.
The operators in Eq.\,\eqref{quadratures} satisfy the following commutation relation:
\begin{equation}\label{quad_comm}
    [\hat a^c, \hat a^s] = i \,.
\end{equation}
{\dr It follows from this commutation relation that if both quadratures are measured with the same precision, then the measurement error is limited by the SQL, which in this particular case has the following form:}
\begin{equation}\label{delta_vac}
  (\Delta a_{\rm SQL})^2 = \frac{1}{2} \,
\end{equation}

Then, we consider two optical modes, the probe and the signal ones, described by the annihilation operators $\hat a$ and $\hat b$, respectively, that interact by means of the {\dr quadratic interaction Hamiltonian of the following form:
\begin{equation}\label{H_QND}
  \dr \mathcal{H} = \hbar\kappa\hat{a}^s\hat{b}^c \,,
\end{equation}
}where $\kappa$ is the interaction strength. In the Heisenberg interaction picture, this Hamiltonian corresponds to the following equations for the quadrature amplitudes of the two modes:
\begin{subequations}\label{QND_quadratures}
    \begin{gather}
    \hat a_{\rm out}^c = \hat a_{\rm in}^c + K\hat b_{\rm in}^c \label{probe}, \\
    \hat b_{\rm out}^c = \hat b_{\rm in}^c, \label{remnant} \\
    \hat a_{\rm out}^s = \hat a_{\rm in}^s, \label{probe_s} \,, \\
    \hat b_{\rm out}^s = \hat b_{\rm in}^s - K\hat a_{\rm in}^s \,,
      \label{signal_s}
    \end{gather}
\end{subequations}
where $K = \kappa\tau$, $\tau$ is the interaction time and the subscripts ``in''  and ``out'' correspond, respectively, to the initial and final state of the system.

The cosine quadrature of the signal mode $\hat b^c$ commutes with the Hamiltonian \eqref{H_QND}, satisfying thus the QND condition. As a result, it remains unchanged during the interaction, see Eq.\,\eqref{remnant}. At the same time, the mode $\hat a$ plays a role of the QND probe: its quadrature $\hat a^c_{\rm out}$ carries out information about $\hat b^c$, which then can be recovered by a subsequent homodyne measurement. It follows from Eq.\,\eqref{probe}, that the corresponding measurement error is equal to
\begin{equation}
  (\Delta b^c_{\rm meas})^2 = \frac{\mean{(\Delta\hat{a}^c_{\rm in})^2}}{K^2} \,,
\end{equation}
where $\mean{(\Delta\hat{a}^c_{\rm in})^2}$ is the initial uncertainty of $\hat{a}^c_{\rm in}$.

Note also that the sine quadrature $\hat b^s$ of the signal mode is perturbed by the probe mode, see Eq.\,\eqref{signal_s}, with the  mean square perturbation equal to
\begin{equation}
  (\Delta b^s_{\rm pert})^2 = K^2\mean{(\Delta\hat{a}^s_{\rm in})^2} \,.
\end{equation}
Taking into account the commutator \eqref{quad_comm}, it is easy to see that $\Delta b^c_{\rm meas}$ and $\Delta b^s_{\rm pert}$ satisfy the Heisenberg uncertainty relation:
\begin{equation}
  (\Delta b^c_{\rm meas})^2(\Delta b^s_{\rm pert})^2
  = \mean{(\Delta\hat{a}^c_{\rm in})^2}\mean{(\Delta\hat{a}^s_{\rm in})^2} \ge \frac{1}{4}
  \,.
\end{equation}

In Ref.\,\cite{yurke1985optical}, a possible implementation of the described above QND scheme was proposed. This scheme is depicted in Fig.\,\ref{fig1}. It uses two DOPAs, each in a separate arm, having opposite signs of the squeeze factor $\pm r$. In order to create the effective interaction between the modes in the form of \eqref{H_QND}, the DOPAs are supplemented by two beamsplitters B1 and B2. The resulting input/output relations are calculated in App.\,\ref{app:A} and have the following form:
\begin{subequations}\label{QND_DOPA2}
    \begin{gather}
        \hat a_{\rm out}^c = -\hat a_{\rm in}^c + 2\hat b_{\rm in}^c\sh r ,\\
        \hat b_{\rm out}^c = \hat b_{\rm in}^c, \label{b_c_inout}\\
        \hat a_{\rm out}^s = -\hat a_{\rm in}^s, \\
        \hat b_{\rm out}^s = \hat b_{\rm in}^s + 2\hat a_{\rm in}^s\sh r ,
    \end{gather}
\end{subequations}
It is easy to see that up to sign flip of $\hat a_{\rm out}^{c,s}$, these equations are equivalent to Eqs.\,\eqref{QND_quadratures} with $K = -2\sh r$.

\section{Measurement schemes based on a single NOPA}\label{sec:schemes}

\subsection{QND measurement}\label{sec:QND}

\begin{figure}
\includegraphics[width=0.7\textwidth]{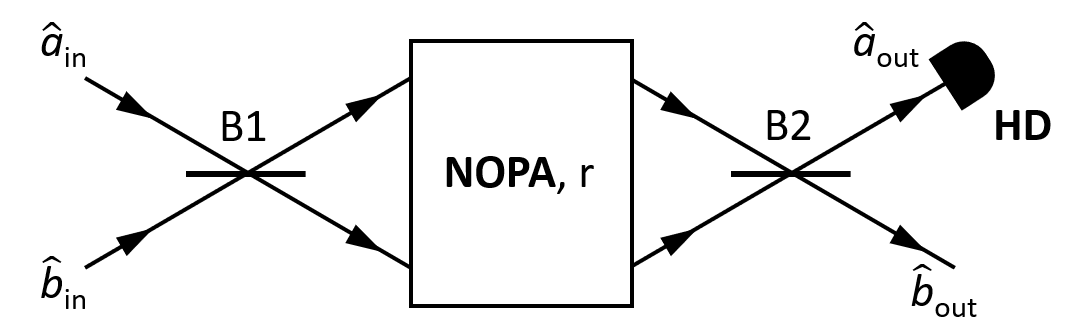}
\caption{\label{fig3} The scheme for the optical field quadrature QND measurement using the non-degenerate optical parametric amplifier (NOPA). ${\rm B}_1$, ${\rm B_2}$ -- beam splitters, HD -- homodyne detector.}
\end{figure}

It is easy to show (see Appendix \ref{B}) that a NOPA with two additional symmetric (50/50) beamsplitters located before and after it, {\dr see Fig.\,\ref{fig:DOPAs},} is equivalent to the pair of DOPAs having the opposite signs of the squeezing factor. These beamsplitters then can be absorbed into the main ones ${\rm B}_{1,2}$, which gives the scheme, shown in Fig.\,\ref{fig3}.


The input-output relations for this scheme read:
\begin{subequations}\label{input-output_NOPA}
  \begin{gather}
    \svector{\hat{a}_{\rm out}^c}{\hat{b}_{\rm out}^c}
      = \mathbb{B}_2\mathbb{S}_N(r)\mathbb{B}_1
          \svector{\hat{a}_{\rm in}^c}{\hat{b}_{\rm in}^c}
      = \smatrix{(T^2-R^2)\ch r}{\sh r + 2RT\ch r}{\sh r - 2RT\ch r}{(T^2 - R^2)\ch r}
          \svector{\hat{a}_{\rm in}^c}{\hat{b}_{\rm in}^c} , \\
    \svector{\hat{a}_{\rm out}^s}{\hat{b}_{\rm out}^s}
      = \mathbb{B}_2\mathbb{S}_N(-r)\mathbb{B}_1
          \svector{\hat{a}_{\rm in}^s}{\hat{b}_{\rm in}^s}
      = \smatrix{(T^2-R^2)\ch r}{-\sh r + 2RT\ch r}{-\sh r - 2RT\ch r}{(T^2 - R^2)\ch r}
          \svector{\hat{a}_{\rm in}^s}{\hat{b}_{\rm in}^s} ,
  \end{gather}
\end{subequations}
where the matrices
\begin{gather}
  \mathbb{S}_N(r) = \smatrix{\ch r}{\sh r}{\sh r}{\ch r} , \\
  \mathbb{B}_1 = \smatrix{-R}{T}{T}{R} \,, \quad \mathbb{B}_2 = \smatrix{R}{T}{T}{-R}
    \label{bbB_12}
\end{gather}
describe the NOPA and the beamsplitters, respectively, with $R$ and $T$ being the amplitude reflectivity and transmissivity coefficients.

Suppose now that
\begin{equation}
  R = \sqrt{\frac{\ch r -1}{2\ch r}} \,,\quad T = \sqrt{\frac{\ch r + 1}{2\ch r}} \,,
\end{equation}
In this case, Eqs.\,\eqref{input-output_NOPA} take the following form:
\begin{subequations}\label{QND_DOPA}
  \begin{gather}
    \hat a_{\rm out}^c = \hat a_{\rm in}^c + 2\hat b_{\rm in}^c\sh r \,, \label{a_inout_c}\\
    \hat b_{\rm out}^c = \hat b_{\rm in}^c,  \label{b_inout_c} \\
    \hat a_{\rm out}^s = \hat a_{\rm in}^s, \\
    \hat b_{\rm out}^s = \hat b_{\rm in}^s - 2\hat a_{\rm in}^s\sh r \,,
  \end{gather}
\end{subequations}
which is identical to Eqs.\,\eqref{QND_quadratures} with $K=2\sh r$.

\subsection{Amplified QND measurement}\label{sec:AQND}

\begin{figure}
\includegraphics[width=0.5\textwidth]{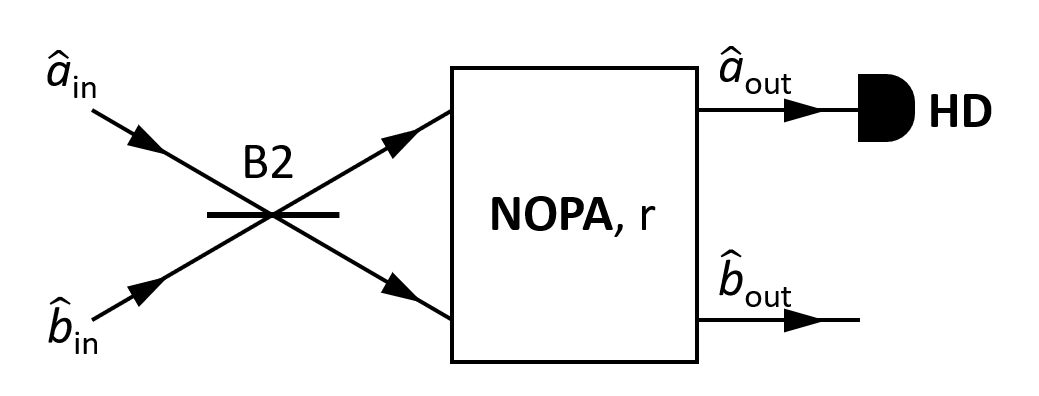}
\caption{\label{fig4} The scheme for the amplified QND measurement of the optical field quadrature. NOPA - nondegenerate optical parametric amplifier, B2 - beamsplitter, HD - homodyne detector.}
\end{figure}

Let us now consider a simpler scheme shown in Fig.\,\ref{fig4}. It consists of a single beamsplitter, followed by the NOPA with the squeeze factor $r$. The input-output relations for this system read:
\begin{subequations}\label{input-output_ampl}
  \begin{gather}
    \svector{\hat{a}_{\rm out}^c}{\hat{b}_{\rm out}^c}
      = \mathbb{S}_N(r){\ds \mathbb{B}_2}\svector{\hat{a}_{\rm in}^c}{\hat{b}_{\rm in}^c}
      = \smatrix{R\ch r + T\sh r}{T\ch r - R\sh r}{T\ch r + R\sh r}{-R\ch r + T\sh r}
          \svector{\hat{a}_{\rm in}^c}{\hat{b}_{\rm in}^c} , \\
    \svector{\hat{a}_{\rm out}^s}{\hat{b}_{\rm out}^s}
      = \mathbb{S}_N(-r){\ds \mathbb{B}_2}\svector{\hat{a}_{\rm in}^s}{\hat{b}_{\rm in}^s}
      = \smatrix{R\ch r - T\sh r}{T\ch r + R\sh r}{T\ch r - R\sh r}{-R\ch r - T\sh r}
          \svector{\hat{a}_{\rm in}^s}{\hat{b}_{\rm in}^s} .
  \end{gather}
\end{subequations}
Choosing the reflectivity and transmissivity coefficients of the beamsplitter as follows:
\begin{equation}
    R = - \frac{1}{A}\ch r, \quad T = \frac{1}{A}\sh r \,,
\end{equation}
where
\begin{equation}
  A=\sqrt{\ch 2r}
\end{equation}
is the amplification factor, we obtain the following set of equations:
\begin{subequations}\label{eq4}
  \begin{gather}
    \hat a^c_{\rm out} = \frac{1}{A}(-\hat a^c_{\rm in} + \hat b_{\rm in}^c\sh 2r), \label{a_inout_2} \\
    \hat b^c_{\rm out} = A\hat b^c_{\rm in} \,, \label{b_inout_2} \\
    \hat a^s_{\rm out} = -A\hat a^s_{\rm in} \,, \\
    \hat b^s_{\rm out} = \frac{1}{A}(\hat b^s_{\rm in} + \hat a^s_{\rm in}\sh 2r) \,,
  \end{gather}
\end{subequations}
compare with Eqs.\,\eqref{QND_DOPA}.

It can be seen from these equations that, unlike the standard QND transformations, the quadratures  $\hat b_{\rm in}^c $ and $\hat a_{\rm in}^s $  are amplified by the factor $A$, which can be modified by changing the NOPA squeezing parameter $r$.

Note also that for the quadrature $\hat{a}^c_{\rm out}$ measured by the homodyne detector, the ratio of the signal term ($\propto \hat{b}^c_{\rm in}$) to the noise term ($\propto \hat{a}^c_{\rm in}$) (the signal-to-noise ratio)  scales with the increase of squeezing as $\sim e^{2r}$, while in the schemes of Fig.\,\ref{fig1} and Fig.\,\ref{fig3} it scales only as $\sim e^r$.

\section{Accounting for the optical losses}\label{sec:losses}

\subsection{Fictional beamsplitters model}\label{sec:FBS}

\begin{figure}
  \includegraphics[width=0.55\textwidth]{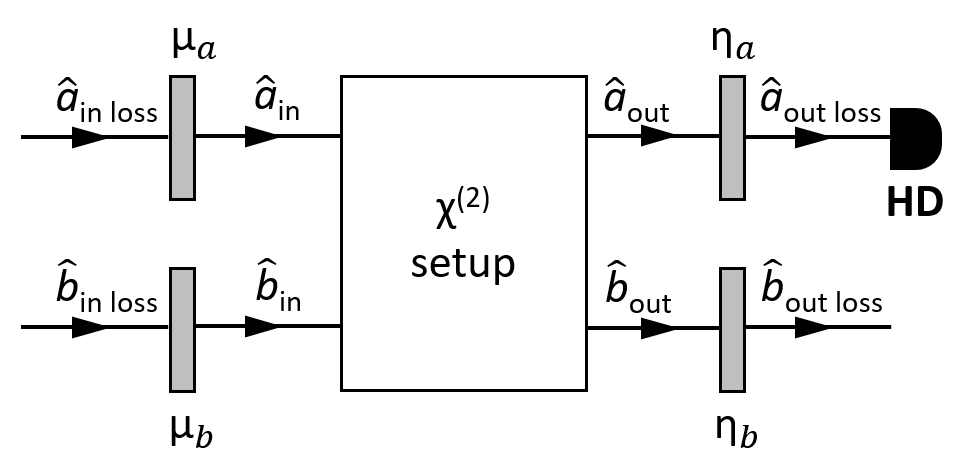}
  \caption{Model of losses based on the fictional beamsplitters. The central block corresponds two the schemes shown in Figs.\,\ref{fig3} and \ref{fig4}.}\label{fig:losses}
\end{figure}

We model the optical losses by means of fictional beamsplitters \cite{yuen1978quantum, leonhardt1995measuring}, which mix the input and output fields of the NOPA with vacuum fields, as it is shown in Fig.\,\ref{fig:losses}. We denote the input fields before the input losses by the subscripts ``in\,loss'', and the output ones after the action of losses ---  by the subscripts ``out\,loss''.

The action of these fictional  beamsplitters is described by the following equations:
\begin{subequations}\label{FBS}
  \begin{gather}
    \hat{a}_{\rm in}^{c,s}
      = \sqrt{\mu_a}\hat{a}_{\rm in\,loss}^{c,s} + \sqrt{1-\mu_a}\hat{u}_a^{c,s} \,,
      \label{a_in_loss} \\
    \hat{b}_{\rm in}^{c,s}
      = \sqrt{\mu_b}\hat{b}_{\rm in\,loss}^{c,s} + \sqrt{1-\mu_b}\hat{u}_b^{c,s} \,,
      \label{b_in_loss} \\
    \hat{a}_{\rm out\,loss}^{c,s}
      = \sqrt{\eta_a}\hat{a}_{\rm out}^{c,s} + \sqrt{1-\eta_a}\hat{v}_a^{c,s} \,,
      \label{a_out_loss} \\
    \hat{b}_{\rm out\,loss}^{c,s}
      = \sqrt{\eta_b}\hat{b}_{\rm out}^{c,s} + \sqrt{1-\eta_b}\hat{v}_b^{c,s} \,,
      \label{b_out_loss}
  \end{gather}
\end{subequations}
where $\mu_{a,b}$ are the quantum  efficiencies of the input ``$a$'' and ``$b$'' channels, $\eta_{a,b}$ are the quantum efficiencies of the output ones, $\hat{u}_{a,b}$ and $\hat{v}_{a,b}$ are vacuum field operators. The relations between {\dr the effective ``lossless''} quadratures $\hat{a}_{\rm in}$, $\hat{b}_{\rm in}$ and $\hat{a}_{\rm out}$, $\hat{b}_{\rm out}$ for the two schemes considered in Sec.\,\ref{sec:QND} and \ref{sec:AQND} are given by Eqs.\,\eqref{QND_DOPA} and \eqref{eq4}, respectively.

Due to the perturbation imposed by the losses, the QND condition \eqref{b_c_inout} and the ``amplified QND'' condition \eqref{b_inout_2} are not exactly fulfilled anymore. Therefore, in the rest of this section, we separately calculate the measurement error for the input quadrature $\hat{b}^c_{\rm in\,loss}$ and the preparation error for the output one $\hat{b}^c_{\rm out\,loss}$.

For simplicity, we assume that the probe beam is initially prepared in the vacuum state. It is easy to see that in this case, $\hat{a}_{\rm in}$ also is a vacuum field. We take into account that the variances of vacuum fields quadratures are equal to the {\dr SQL \eqref{delta_vac}}.

\subsection{Measurement}

First, we consider the problem of measuring the quadrature $\hat b^c_{\rm in}$. For this we assume a homodyne measurement of the quadrature $\hat a^c_{\rm out\,loss}$. In the case of the QND measurement scheme of Sec.\,\ref{sec:QND}, combining Eqs.\,\eqref{a_inout_c}, \eqref{b_in_loss}, and \eqref{a_out_loss}, we obtain that
\begin{equation}\label{a_out_meas}
  \hat{a}_{\rm out\,loss}^c = G_{\rm meas}\hat{b}_{\rm in\,loss}^c + \hat{w}_{\rm meas} \,,
\end{equation}
where
\begin{equation}
  G_{\rm meas} = 2\sqrt{\eta_a\mu_b}\sh r
\end{equation}
is the gain factor and
\begin{equation}
  \hat{w}^c_{\rm meas}
  = \sqrt{\eta}_a
      \bigl(a^c_{\rm in} + 2\sqrt{1-\mu_b}\hat{u}_b^c\sh r\bigr) + \sqrt{1-\eta_a}\hat v_a^c
\end{equation}
is the total measurement noise. Therefore, the measurement error for $\hat{b}^c_{\rm in\,loss}$ is equal to
\begin{equation}\label{meas_QND}
  (\Delta b^c_{\rm meas})^2
  = \frac{\mean{(\hat{w}_{\rm meas}^c)^2}}{G_{\rm meas}^2}
  = (\Delta b^c_{\rm meas\,0})^2 + \frac{1}{8\mu_b\eta_a\sh^2r} \,,
\end{equation}
where
\begin{equation}\label{d_b_meas_0}
  (\Delta b^c_{\rm meas\,0})^2 = \frac{1-\mu_b}{2\mu_b}
\end{equation}
is the part imposed by losses in the input path of the signal mode. This part does not depend on the measurement procedure.

In similar way, in the case of the amplified QND measurement of Sec.\,\ref{sec:AQND}, combining Eqs.\,\eqref{a_inout_2}, \eqref{b_in_loss}, and \eqref{a_out_loss}, we again come to Eq.\,\eqref{a_out_meas}, but with the gain factor and the total noise equal to
\begin{equation}
  G_{\rm meas} = \frac{\sqrt{\eta_a\mu_b}\sh2r}{A}
\end{equation}
and
\begin{equation}
  \hat{w}^c_{\rm meas}
  = \frac{\sqrt{\eta}_a}{A}(-\hat a^c_{\rm in} + \sqrt{1-\mu_b}\hat{u}_b^c\sh2r)
    + \sqrt{1-\eta_a}\hat v_a^c \,.
\end{equation}
Therefore, in this case, the measurement error for $\hat{b}^c_{\rm in\,loss}$ is equal to
\begin{equation}\label{meas_AQND}
  (\Delta b^c_{\rm meas})^2 = \frac{\mean{(\hat{w}_{\rm meas}^c)^2}}{G_{\rm meas}^2}
  = (\Delta b^c_{\rm meas\,0})^2
    +  \frac{1}{2\mu_b\sh^22r}\biggl(\frac{1-\eta_a}{\eta_a}\ch2r + 1\biggr) .
\end{equation}

\subsection{Preparation}

In the presence of losses, the input and output values of the measured quadrature are not equal to each other: $\hat{b}_{\rm in\,loss}\ne\hat{b}_{\rm out\,loss}$. Therefore, a data processing procedure that differs from that for the input value measurement is required for the optimal output value estimate.

It should also be noted that, in general, the quantum state preparation result depends on the {\it a priori} information on the measured value. However, the goal of this work is to explore performance of the preparation procedure alone. Therefore, we assume here that no information about the initial state of the signal mode is available.

In the case of the QND measurement scheme of Sec.\,\ref{sec:QND}, combining Eqs.\,\eqref{a_inout_c}, \eqref{b_inout_c}, \eqref{a_out_loss}, and \eqref{b_out_loss}, we obtain
\begin{equation}\label{a_out_prep}
  \hat{a}_{\rm out\,loss}^c = G_{\rm prep}\hat{b}_{\rm out\,loss}^c + \hat{w}_{\rm prep}^c \,,
\end{equation}
where
\begin{equation}
  G_{\rm prep} = 2\sqrt{\frac{\eta_a}{\eta_b}}\sh r
\end{equation}
is the gain factor and
\begin{equation}
  \hat{w}^c_{\rm prep}
  = \sqrt{\eta}_a
      \biggl(a^c_{\rm in} - 2\sqrt{\frac{1-\eta_b}{\eta_b}}\hat{v}_b^c\sh r\biggr)
    + \sqrt{1-\eta_a}\hat v_a^c
\end{equation}
is the total noise. Therefore, the estimation uncertainty for $\hat{b}^c_{\rm out\,loss}$ (the preparation error) is equal to
\begin{equation}\label{prep_QND}
  (\Delta b^c_{\rm prep})^2
  = \frac{\mean{(\hat{w}_{\rm prep}^c)^2}}{G_{\rm prep}^2}
  = (\Delta b^c_{\rm prep\,0})^2 + \frac{\eta_b}{8\eta_a\sh^2r} .
\end{equation}
where
\begin{equation}\label{d_b_prep_0}
  (\Delta b^c_{\rm prep\,0})^2 = \frac{1-\eta_b}{2}
\end{equation}
is the part imposed by the losses in the output path of the signal mode.

For the amplified QND measurement of Sec.\,\ref{sec:AQND},  combining Eqs.\,\eqref{a_inout_2}, \eqref{b_inout_2}, \eqref{a_out_loss}, and \eqref{b_out_loss}, we obtain Eq.\,\eqref{a_out_prep} with the gain factor and the sum noise equal to, respectively,
\begin{equation}\label{G_prep_2}
  G_{\rm prep} = \sqrt{\frac{\eta_a}{\eta_b}}\,\tanh 2r
\end{equation}
and
\begin{equation}
  \hat{w}^c_{\rm prep}
  = -\sqrt{\eta}_a\biggl(
        \frac{a^c_{\rm in\,0}}{A}
        + \sqrt{\frac{1-\eta_b}{\eta_b}}\frac{\sh2r}{A^2}\hat{v}_b^c
      \biggr)
    + \sqrt{1-\eta_a}\hat v_a^c \,,
\end{equation}
which corresponds to the preparation error
\begin{equation}\label{prep_AQND}
  (\Delta b^c_{\rm prep})^2
  = \frac{\mean{(\hat{w}_{\rm prep}^c)^2}}{G_{\rm prep}^2}
  = (\Delta b^c_{\rm prep\,0})^2
    + \frac{\eta_b}{2\tanh^22r}\biggl(\frac{1}{\cosh2r} + \frac{1-\eta_a}{\eta_a}\biggr) .
\end{equation}

\subsection{Estimates}

All four equations \eqref{meas_QND}, \eqref{meas_AQND}, \eqref{prep_QND}, and \eqref{prep_AQND} obtained in this section have the same structure. They consist of two terms. The first ones, $(\Delta b^c_{\rm meas\,0})^2$ in measurement case and $(\Delta b^c_{\rm prep\,0})^2$ in the preparation, originate from the losses in the input and output paths of the signal mode, respectively. {\dr Therefore, they} are invariant to the specific measurement/preparation procedure, and do not depend on the squeeze factor $r$. The second terms represent the measurement/preparation errors proper. {\dr In all these cases, the errors sharply decrease with increase of $r$. In the QND measurement and preparation cases, as well as in the AQND measurement case, the sensitivity is limited by the respective asymptotic values \eqref{d_b_meas_0} and \eqref{d_b_prep_0}.

However, this is not the case for the AQND preparation due to the additional term $(1-\eta_a)/\eta_a$ in Eq.\,\eqref{prep_AQND}. This asymptotically independent on $r$ term appears because the gain factor \eqref{G_prep_2} corresponds to the {\it already amplified} by $A\propto e^r$ value of the measured quadrature $b^c$ and therefore does not depend on $r$ at $r\gg1$.

For our estimates, we use the values of squeezing up to $\sim20\,{\rm dB}$.} For the quantum efficiency of all input/output channels, with the exception of the detection efficiency, we assume the reasonably optimistic values of
\begin{equation}
  \mu_a = \mu_b = \eta_b = 0.95.
\end{equation}
In the detection channel, we take into account the quantum efficiency of the homodyne detector, assuming that
\begin{equation}
  \eta_a = 0.9
\end{equation}

\begin{figure}
  \includegraphics[width=0.7\textwidth]{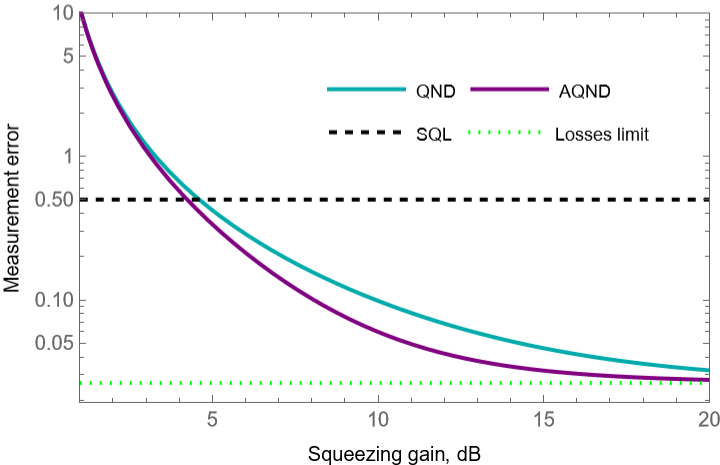}
  \caption{Measurement errors \eqref{meas_QND} and \eqref{meas_AQND} as functions of the squeezing gain $e^r$ expressed in dB. For the comparison, the vacuum state uncertainty $\Delta_{\rm vac}$ (SQL) and the asymptotic limit \eqref{d_b_meas_0} are also presented.} \label{fig:meas}
\end{figure}

\begin{figure}
  \includegraphics[width=0.7\textwidth]{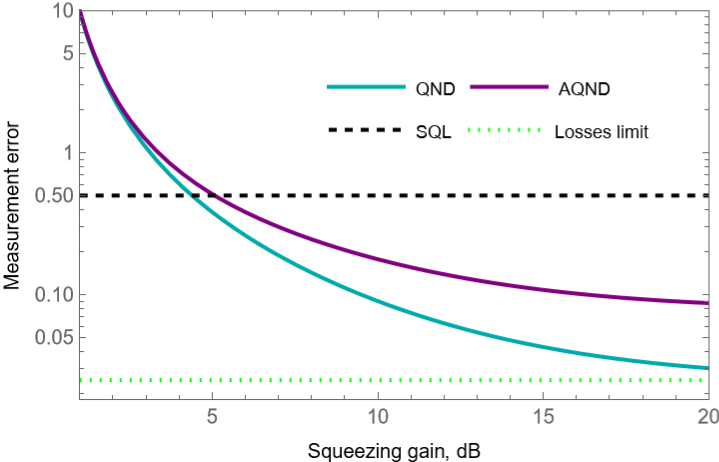}
  \caption{Measurement errors \eqref{meas_QND} and \eqref{meas_AQND} as functions of $e^r$ expressed in dB. For the comparison, the vacuum state uncertainty $\Delta_{\rm vac}$ (SQL) and the asymptotic limit \eqref{d_b_prep_0} are also presented.} \label{fig:prep}
\end{figure}

{\dr In Figs.\,\ref{fig:meas} and \ref{fig:prep}, the corresponding measurement and preparation errors are plotted as functions of $r$. It can be seen from these plots, that for the reasonably optimistic values of squeezing, the sensitivity several times better than the SQL can be achieved.}

\section{Discussion}\label{III}

{\dr

The QND measurements of optical quadratures open new possibilities for optimizing quantum optical measurement strategies. They could also be of significant interest for quantum information processing, particularly for continuous-variable quantum computing.

In this work, we proposed a quantum QND scheme for measuring quadrature amplitudes based on a single non-degenerate optical parametric amplifier supplemented by two passive optical elements (beamsplitters). We further developed this scheme to a novel type of QND measurement that allows for amplification of the measured quadrature, but still does not introduce any back-action noise into it. Such amplification may find interesting applications in systems where the signal is attenuated, for example due to losses.

The schemes discussed here have a general structure and can be implemented on various platforms using current experimental achievements. The key requirement for these platforms is the sufficient degree of nonlinear interaction, corresponding to $\gtrsim10\,{\rm db}$ of squeezing.

Using high concentration of the optical power in short (picosecond) pump pulses, significant squeezing can be achieved in a single cavity-less nonlinear crystals. For example, in the work \cite{frascella2021overcoming}, 15\,dB of initial (before the losses) squeezing and 31\,dB of parametric amplification was demonstrated using BBO crystals.  In Ref.\,\cite{Barakat25}, 5.2\,dB of squeezing with an account for the 15\% losses was achieved.

Photonic waveguides allow for increasing the length of the nonlinear interaction and thus achieving a strong squeezing. In particular, 6.3 dB of squeezing was achieved in \cite{10.1063/5.0063118} using a periodically poled lithium niobate (PPLN) waveguide, while 4.9 dB of squeezing was reached over a broad optical range in \cite{doi:10.1126/science.abo6213}. In the work \cite{jankowski2022quasi}, the parametric gain (the anti-squeezing) as large as 71\,dB was obtained using the periodically poled thin-film lithium niobate nanowaveguide.

By placing the nonlinear crystals into optical cavities, significant degree of squeezing can be obtained in the continuous-wave regime. For example, the squeezing of -15 dB was achieved using the periodically poled titanil phosphate (PPKTP) crystal assisted by an optical cavity \cite{PhysRevLett.117.110801}.

For the quantum information processing tasks, the most promising platform is probably optical microresonators. In \cite{Zhao_PRL_124_193601_2020},  -3.09 dB  of quadrature squeezing on chip was demonstrated by employing a degenerate four-wave-mixing  process in integrated silicon-nitride microresonator. In \cite{Ulanov} the photonic crystal microresonator had a tailored nano-corrugation, that modified resonances to suppress parasitic nonlinear processes, resulting in 7.8 dB of on-chip squeezing.

With an account for these results, the practical implementation of the QND schemes discussed in this paper can be considered feasible.

}

\acknowledgments

The work of D.S. and F.K. was supported by the Russian Science Foundation (Project No. 25-12-00263).

\appendix

\section{Analysis of DOPA based QND scheme}\label{app:A}

The input-output relations of all three components of the scheme shown in Fig.\,\ref{fig1} are the following:
\begin{subequations}\label{input-output_DOPA}
  \begin{gather}
    \svector{\hat{b}_{\rm out}^c}{\hat{a}_{\rm out}^c}
      = \mathbb{B}_2\mathbb{S}_D(r)\mathbb{B}_1
          \svector{\hat{a}_{\rm in}^c}{\hat{b}_{\rm in}^c}
      = \smatrix{T^2e^{-r} - R^2e^r}{2RT\ch r}{-2RT\ch r}{T^2e^r-R^2e^{-r}}
          \svector{\hat{a}_{\rm in}^c}{\hat{b}_{\rm in}^c} , \\
    \svector{\hat{b}_{\rm out}^s}{\hat{a}_{\rm out}^s}
      = \mathbb{B}_2\mathbb{S}_D(-r)\mathbb{B}_1
          \svector{\hat{a}_{\rm in}^s}{\hat{b}_{\rm in}^s}
      = \smatrix{T^2e^{r} - R^2e^{-r}}{2RT\cosh r}{-2RT\cosh r}{T^2e^{-r}-R^2e^{r}}
          \svector{\hat{a}_{\rm in}^s}{\hat{b}_{\rm in}^s} ,
  \end{gather}
\end{subequations}
where the matrices $\mathbb{S}_D$ and $\mathbb{B}_{1,2}$ are given by Eqs.\,\eqref{bbS_D}, \eqref{bbB_12}, respectively. Note that the signal and probe output fields are swapped.

If
\begin{equation}
    T = \frac{e^{r}}{\sqrt{1+e^{2r}}} \,, \quad
    R = \frac{1}{\sqrt{1+e^{2r}}} ,
\end{equation}
then Eqs.\,\eqref{input-output_DOPA} reduce to Eqs.\,\eqref{QND_DOPA2}.

{\ds \section{Equivalence of NOPA and two antisymmetric DOPAs}\label{B}

\begin{figure}
\includegraphics[width=\textwidth]{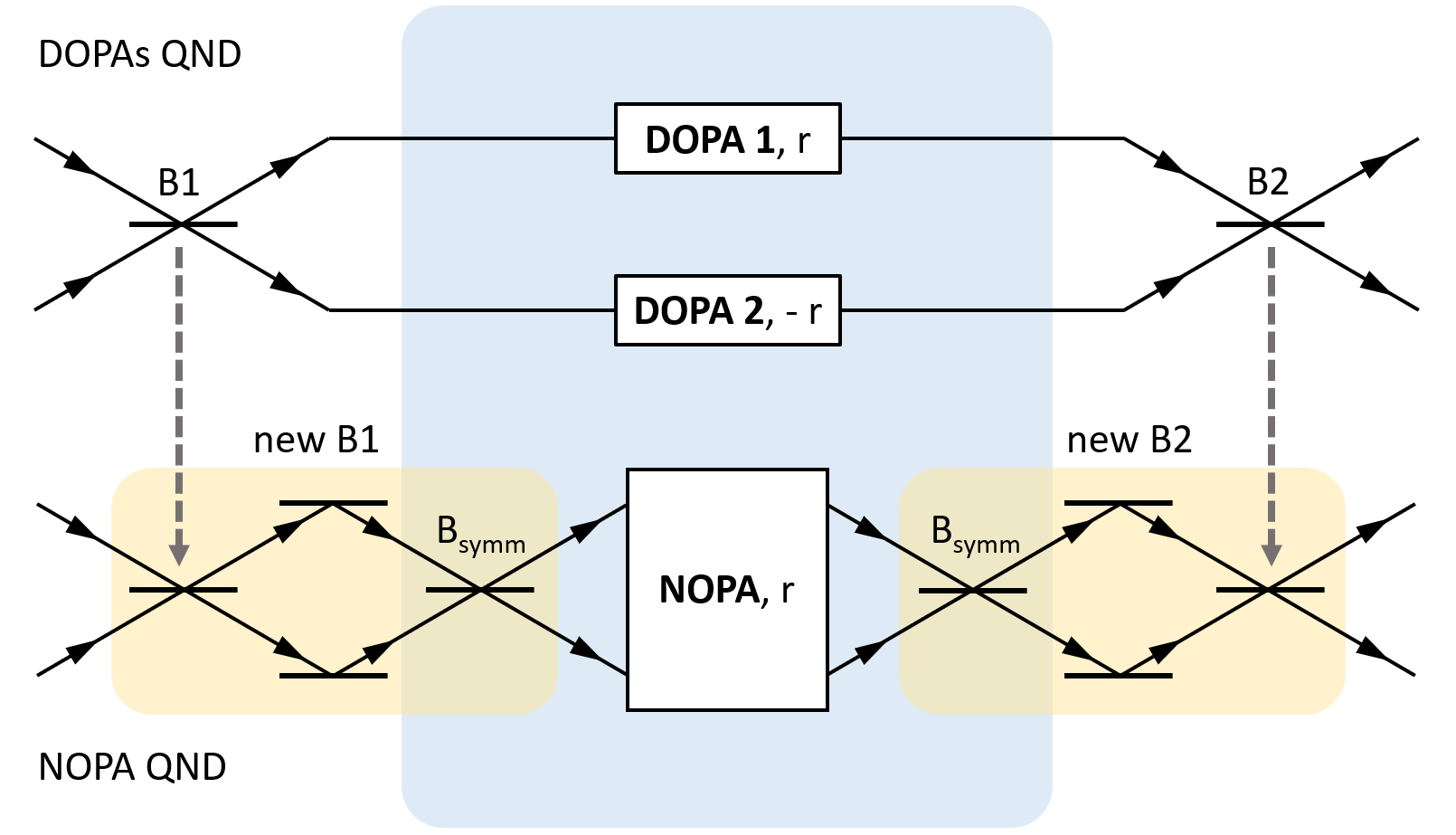}
\caption{\label{fig5} {\ds Equivalence of the DOPAs and NOPA based schemes. Blue area represents equivalent parts in both discussed schemes. Each yellow area highlights two beamsplitters which can be transformed into one with adjusted parameters. DOPA -- degenerate optical parametric amplifier, NOPA -- nondegenerate optical parametric amplifier, ${\rm B}_1$, ${\rm B_2}$ -- beam splitters, HD -- homodyne detection.}
}
\end{figure}

Consider two schemes shown in Fig.\,\ref{fig5}

The action of a NOPA can be described by the following equations:
\begin{equation}
  \svector{\hat{a}_{2}^c}{\hat{b}_{2}^c}
    = \mathbb{S}_N(r)\svector{\hat{a}_1^c}{\hat{b}_1^c} , \quad
  \svector{\hat{a}_{2}^s}{\hat{b}_{2}^s}
    = \mathbb{S}_N(-r)\svector{\hat{a}_1^s}{\hat{b}_1^s} ,
\end{equation}
while the action of the symmetric beamsplitter reads:
\begin{equation}
  \svector{\hat{a}_{2}^{c,s}}{\hat{b}_{2}^{c,s}}
    = \mathbb{B}_{\rm symm}\svector{\hat{a}_1^{c,s}}{\hat{b}_1^{c,s}} \,,
\end{equation}
where $\hat{a}_{1}^{c,s}$, $\hat{b}_{1}^{c,s}$ and  $\hat{a}_{2}^{c,s}$, $\hat{b}_{2}^{c,s}$ are the quadratures of the input and output fields, respectively, and
\begin{equation}
  \mathbb{B}_{\rm symm} = \frac{1}{\sqrt{2}}\smatrix{1}{1}{1}{-1} ,
\end{equation}
It is easy to see that
\begin{equation}
  \mathbb{B}_{\rm symm}\mathbb{S}_N(\pm r)\mathbb{B}_{\rm symm} = \mathbb{S}_D(\pm r) \,,
\end{equation}
where the matrix
\begin{equation}\label{bbS_D}
  \mathbb{S}_D(r) = \smatrix{e^r}{0}{0}{e^{-r}} \,
\end{equation}
describes the DOPA.


\providecommand{\noopsort}[1]{}\providecommand{\singleletter}[1]{#1}%

\end{document}